\documentclass{sigchi}

\usepackage{balance}  
\usepackage{graphics} 
\usepackage{txfonts}
\usepackage{times}    
\usepackage[pdftex]{hyperref}
\usepackage{color}
\usepackage{textcomp}
\usepackage{booktabs}
\usepackage{ccicons}
\usepackage{todonotes}

\usepackage{comment}
\usepackage{times,listings,amsfonts}
\usepackage{amsmath,xcolor}
\usepackage{pifont}
\usepackage{xspace}
\usepackage{url}
\usepackage{multirow}

\newcommand{\nMonth}{21 }
\newcommand{\nSurvey}{32 }
\newcommand{\nInterview}{9 }
\newcommand{\nUser}{220 }

\makeatletter
\def\url@leostyle{%
  \@ifundefined{selectfont}{\def\UrlFont{\sf}}{\def\UrlFont{\small\bf\ttfamily}}}
\makeatother
\def\pprw{8.5in}
\def\pprh{11in}

\definecolor{linkColor}{RGB}{6,125,233}
\hypersetup{%
  pdftitle={SIGCHI Conference Proceedings Format},
  pdfauthor={LaTeX},
  pdfkeywords={SIGCHI, proceedings, archival format},
  bookmarksnumbered,
  pdfstartview={FitH},
  colorlinks,
  citecolor=black,
  filecolor=black,
  linkcolor=black,
  urlcolor=linkColor,
  breaklinks=true,
}

\begin{document}

\title{Genie: A Longitudinal Study Comparing Physical and\\[0.1cm] Software-augmented Thermostats in Office Buildings\\[-0.5cm]}

\numberofauthors{1}
\author{
  \alignauthor Bharathan Balaji$^\dag$, Jason Koh$^\dag$, Nadir Weibel$^\dag$, Yuvraj Agarwal$^\ddag$\\ 
  \vspace{2mm} 
  \affaddr{$^\dag$University of California, San Diego \hspace{12mm} $^\ddag$Carnegie Mellon University}\\
  \email{$^\dag$\{bbalaji, jbkoh, weibel\}@ucsd.edu, \hspace{5mm}$^\ddag$yuvraj.agarwal@cs.cmu.edu}
}

\maketitle

\begin{abstract}

Thermostats are primary interfaces for occupants of office buildings to express their comfort preferences. However, standard thermostats are often ineffective due to inaccessibility, lack of information, or limited responsiveness, leading to occupant discomfort. Software thermostats based on web or smartphone applications provide alternative interfaces to occupants with minimal deployment cost. However, their usage and effectiveness have not been studied extensively in real settings. In this paper we present \emph{Genie}, a novel software-augmented thermostat that we deployed and studied at our university over a period of \nMonth months. Our data shows that providing wider thermal control to users does not lead to system abuse and that the effect on energy consumption is minimal -- while improving comfort and energy awareness. We believe that increased introduction of software thermostats in office buildings will have important effects on comfort and energy consumption and we provide key design recommendations for their implementation and deployment.

 
\end{abstract}


\category{H.5.3.}{Information Interfaces and Presentation
  (e.g. HCI)}{Group and Organization Interfaces, Web-based interaction}

\section{Introduction}
Office buildings' occupants interact with the HVAC system (Heating, Ventilation and Air Conditioning) using thermostats which provide information such as current room temperature and whether HVAC is operating, as well as enable minor adjustments to the temperature settings. Since the ability to maintain control over their thermal environment has been shown to have a major effect on occupant satisfaction~\cite{dear2013progress,paciuk1989role}, it is critical that these devices are accurate, effective and usable by occupants. In addition, thermostats are a key component of HVAC operation as they complete the feedback loop in the control system and provide insights into several types of HVAC faults.

Most buildings typically use a variant of the ubiquitous \textit {physical thermostat}, under the assumption that they are intuitive to use, without any occupant training. However, a recent survey of 215 buildings across US, Canada and Finland showed that 89\% of the buildings do not meet thermal comfort standards~\cite{huizenga2006air}. More importantly, in the survey three of the top five reasons linked to occupant dissatisfaction were due to thermostats, specifically (a) thermostats are inaccessible, (b) thermostats are controlled by other people, and (c) HVAC systems do not respond quickly enough to changes on the thermostat. Meier et al.~\cite{meier2011thermostat} studied the various thermostat designs available today and confirmed how a poor user interface (UI) and occupants' misconceptions have a significant impact on comfort and HVAC energy consumption. 



\emph{Software thermostats} provide an attractive alternative to physical thermostats~\cite{balaji2013zonepac,erickson2012thermovote,jazizadeh2013human}. They provide occupants with an interface to the HVAC system via a web service or a native application, allowing them to have personalized settings that maximize comfort. Erickson et al~\cite{erickson2012thermovote} showed that use of a native application feedback system led to an improvement in user satisfaction from 25\% to 100\% in a university building. Furthermore, unlike physical thermostats, software thermostats are incrementally deployable within existing HVAC systems, and are continuously upgradeable with new features or updates to control policies. 


In order to investigate the usage of software thermostats and their impact on comfort and energy consumption, we designed and deployed \emph{Genie}, a software-augmented thermostat, directly integrated with our building's HVAC system. Genie displays all essential information conveyed by traditional thermostats in a web application. Since software interfaces can be made richer than physical thermostats, Genie supports additional features such as (i) the ability for occupants to send thermal feedback to building managers, (ii) the display of current weather conditions, (iii) an expanded level of control of the local temperature to $\pm3^{\circ}$F, and (iv) the ability to turn On/Off HVAC as needed. Additionally, Genie estimates the energy use by each thermal zone using heat transfer equations~\cite{balaji2013zonepac} and display the results to the occupants of that space as a way to measure their energy impact.  

To study real world usage of Genie, we deployed it in our university building (approximately 150,000 sqft., five floors). Genie has been in use by \nUser users over the period of \nMonth months and in this paper we present a detailed analysis of its usage. We further augment our analysis with survey and interviews conducted at the end of our study to assess the usefulness and usability of Genie to the building occupants. As far as we are aware, this is the first longitudinal study of physical and software office thermostats at a large scale. 

\subsection{Contributions}
 Our data show several interesting findings that can serve as key design recommendations for implementation and deployment of software thermostats. We observe that the majority of thermostats are seldom used and find that some thermostats change temperature settings erroneously without user input, leading to significant discomfort and equipment damage. Additionally, our paper confirms observations made in prior work that occupants have misconceptions about thermostat operation, and resort to improvisations when they are uncomfortable. Our data also shows that occupants are more comfortable with additional status information and added control for the HVAC system, and that electronic occupants' feedback about their comfort provided immediate insights into HVAC's usage characteristics and faults.
 
 All in all, the study we present here indicates how providing wider thermal control to users does not lead to system abuse and the effect on energy consumption is minimal -– while improving comfort and energy awareness.

\section{Background and Related Work}
\label{sec:background}
Maintaining occupant thermal comfort is essential for a satisfactory~\cite{fountain1996expectations} and productive~\cite{seppanen2006room} office environment, and studies show that effective HVAC control by occupants themselves is key~\cite{dear2013progress,paciuk1989role,van2008forty}. Hence, thermostats and thermal comfort have been studied extensively ~\cite{dear2013progress,karjalainen2007user,peffer2011people,van2008forty,van2010thermal}. The usability of residential thermostats has been explored in depth~\cite{peffer2011people}, where thermostats have evolved from simple mechanical devices to digital programmable thermostats. The latest devices even include network connectivity, learning, energy feedback and updated UIs for occupant interaction\footnote{Nest: https://nest.com/, Ecobee: https://www.ecobee.com/}. On the other hand, the long-term usage of thermostats in office buildings has not been studied as much. 


The thermal comfort model followed in most buildings in the US is specified by ASHRAE Standard 55~\cite{standard2004standard}, itself based on Fanger's Predicted Mean Vote (PMV) model~\cite{fanger1970thermal}. Fanger's PMV model considers various parameters such as air temperature, air velocity, humidity, clothing insulation and metabolism of the occupant to predict occupant comfort. The PMV expresses comfort with a 7-point scale, ranging from Hot(+3) to Cold(-3), and occupants are considered comfortable if the PMV is between +1 and -1. Using this model, engineers design systems to maintain a range of temperature that satisfies at least 80\% of the occupants, and provide local control options for minor changes to the temperature setting. 

Several studies have shown that occupants are not comfortable in office spaces~\cite{bordass1993user,huizenga2006air,karjalainen2009thermal,karjalainen2007user,owners1800managers}. A survey by Huizenga et al.~\cite{huizenga2006air} shows that 89\% of buildings do not meet comfort standards and lists (a) hot/cold regions, (b) thermostat inaccessibility and (c) thermostats controlled by other people, as primary reasons for discomfort. Contextual interviews by Karjalainen et al.~\cite{karjalainen2007user} found that users are unaware that thermostat exists, thermostats are inaccessible, they lack informative feedback, users think they are not allowed to control the thermostat, thermostat's dial is stiff or broken, and -- most commonly -- users did not know how much the thermostat dial should be turned to get desired room temperature. In a follow up work, Karjalainen et al.~\cite{karjalainen2010usability} provide design guidelines based on user studies for office thermostats emphasizing clarity of information, adequate control, acceptable default settings, informative help and aesthetics. However, these guidelines were not tested in practice. 

Several variations of software thermostats have been proposed to improve the interaction between occupants and the HVAC system. Murakami et al.~\cite{murakami2007field} introduced a desktop voting system that determines the temperature of the entire floor based on occupants' feedback. Occupants provide feedback whether they want temperature to be warmer or colder, and communicate comfort level on the standard 7-point scale. Although the system showed a promising 20\% energy savings, it was only deployed for a few days. Jazizadeh et al.~\cite{jazizadeh2013human} developed a smartphone application that lets occupants provide feedback on required temperature, airflow and lighting level. Their input is mapped to a learning model to determine the HVAC settings. However, they do not deploy their system for real use. Thermovote~\cite{erickson2012thermovote} seeks to overcome the limitations of the PMV model by using a software interface to gather occupants' comfort levels in the standard 7-point scale. The occupant feedback was used to estimate a corrected PMV and the temperature settings of the office are adjusted automatically. User satisfaction rose from 25\% to 100\% with this strategy over a period of 5 months, with a decrease of 10\% in energy consumption. However, the occupants were prompted every 10 minutes for their comfort feedback and were not provided any other feedback on the current status of HVAC. Comfy\footnote{https://gocomfy.com/} provides a web interface to office occupants to collect their comfort feedback. The occupants are given a choice between ``Warm'' and ``Cold'', and their feedback is used to adjust the temperature setting for the room. These temperature settings are gradually relaxed over time until there is another occupant input from the web interface. Occupants are provided no other information than the simplified ``Warm'' and ``Cool'' buttons. Comfy's case study reports engagement of 77\% of the users across 6 months and an energy reduction of 22\% due to the relaxed setting employed when there is no input from occupants. 

These prior work show the promise of software thermostats to overcome limitations of physical thermostat  controls. However, these systems also force users to engage with the system while providing no information on the current HVAC status. It is also unclear how the existing thermostat works with these software systems and what happens when users do not have access to a computer or when there is a software failure. No user study has been conducted to investigate these aspects. Furthermore, the onus of maintenance of these systems is on the building manager, and prior studies indicate that building managers are already overwhelmed with HVAC management issues~\cite{mills2011building,teraoka2014buildingsherlock}. 

We propose an alternative design approach where occupants are provided with essential information such as current room temperature and setpoints, allowing them to take control of their environment and send feedback based on the information provided. Balaji et al.~\cite{balaji2013zonepac} designed a web application that shows the HVAC system status, allows occupants to control their settings and send comfort feedback. This work focused on providing accurate per-zone energy feedback and on quantifying the effect on energy consumption when using a software thermostat prototype across five days. We use a similar design strategy, but study the effect of usage across \nMonth months. To the best of our knowledge, none of the prior work has studied the actual use of physical or software thermostats in a longitudnal study at a large scale. We compare the usage of hardware thermostats with Genie, studying their use in isolation and when combined. We also show how users' feedback can be valuable in fault identification, and how information about energy usage improves overall awareness.

\section{University Building Testbed}
\begin{figure}[t!]
\includegraphics[width=\linewidth]{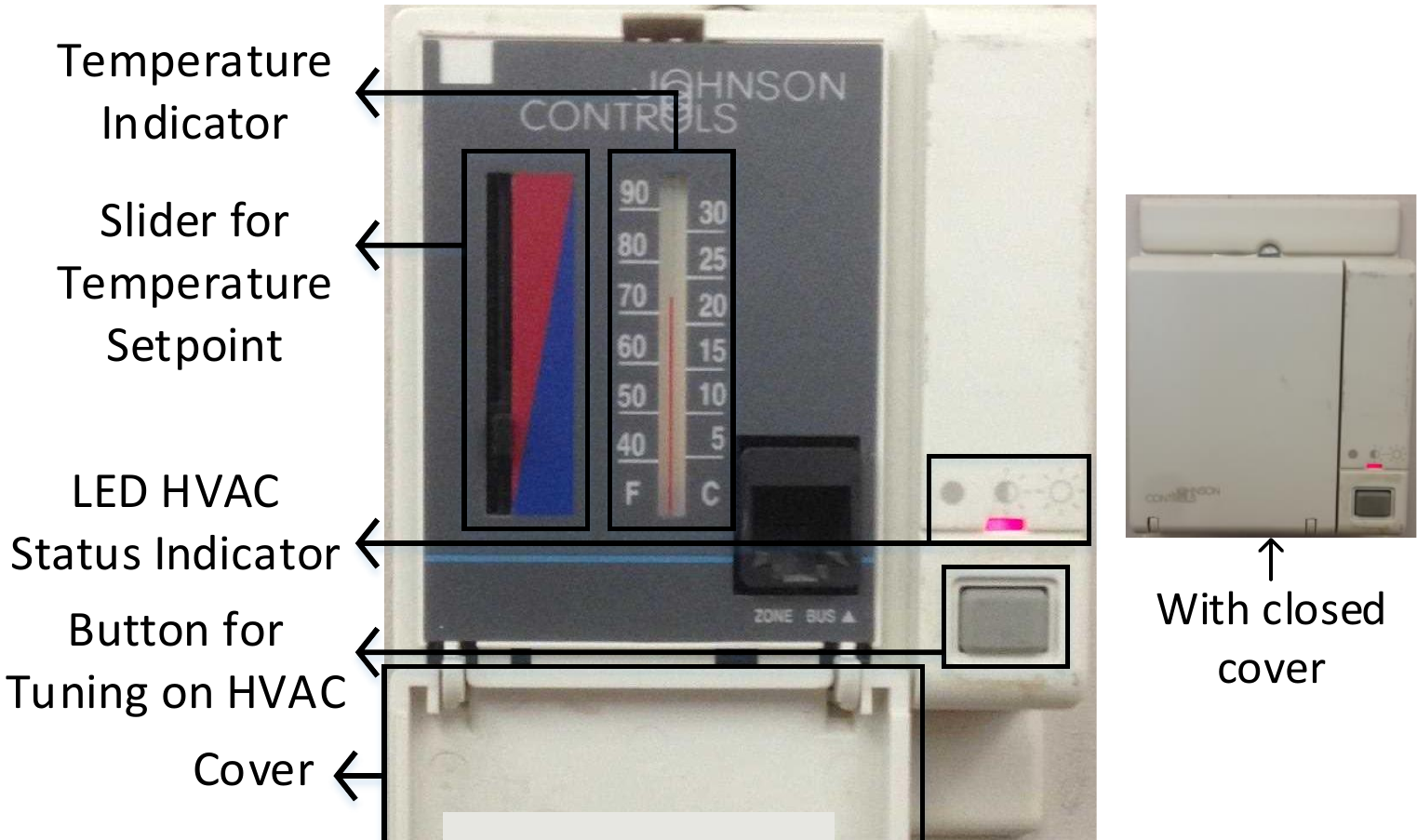}
\caption{Thermostat used in the CSE building. Slider adjusts temperature setpoint by $\pm1^oF$. HVAC power button turns On HVAC for 2 hours on nights/weekends.}
\label{fig:cse_thermostat}
\vspace{-5mm}
\end{figure}

We use a 150,000 sq-ft. university building consisting of five floors and 236 \emph{thermal zones} as our testbed. Each thermal zone typically consists of a large room such as a conference room or multiple small offices. In both cases HVAC is managed by a single thermostat. Figure \ref{fig:cse_thermostat} shows the annotated picture of the thermostat in use in our building. 

From Figure \ref{fig:cse_thermostat} we can see that when the thermostat cover is closed, its functionality is somewhat unclear to occupants. Once open, the thermostat consists of an analog thermometer and a slider to adjust the temperature setpoint by $\pm1^{\circ}$F. However, since there is no quantitative feedback on the effect of adjusting the slider, occupants are often unsure about its effect. In reality, the change in temperature due to the slider position is often non-linear and differs betweens zones depending on the degree of flexibility provided by the building manager in response to comfort complaints. Thus, occupants experience is inconsistent across different thermostats.

The LED on the panel indicates system status for that zone -- when the LED is On (pink) the HVAC is in \emph{Occupied} mode, when blinking it is in \emph{Stand-by} mode and if the LED is Off, the HVAC is in \emph{Unoccupied} mode. In the Occupied mode, the room temperature is kept within a 4$^{\circ}$F bound with adequate airflow; in the Standby and Unoccupied mode the temperature band changes to 8$^{\circ}$F and 12$^{\circ}$F respectively with minimal airflow. The HVAC system runs on a static schedule: 6am to 6pm in Occupied mode, 6pm to 10pm in Standby on weekdays, and in Unoccupied mode for nights and weekends. If the occupants are in the building during off hours, they are expected to push the grey button to put the system into the Occupied mode for 2 hours. From Fig.~\ref{fig:cse_thermostat}, we can see that these features are not apparent without prior knowledge. 

%
%
%
%
%
%
%
%
%
%

\subsection{Thermal Zone's Temperature Control}
\begin{figure}[t!]
\includegraphics[width=\linewidth]{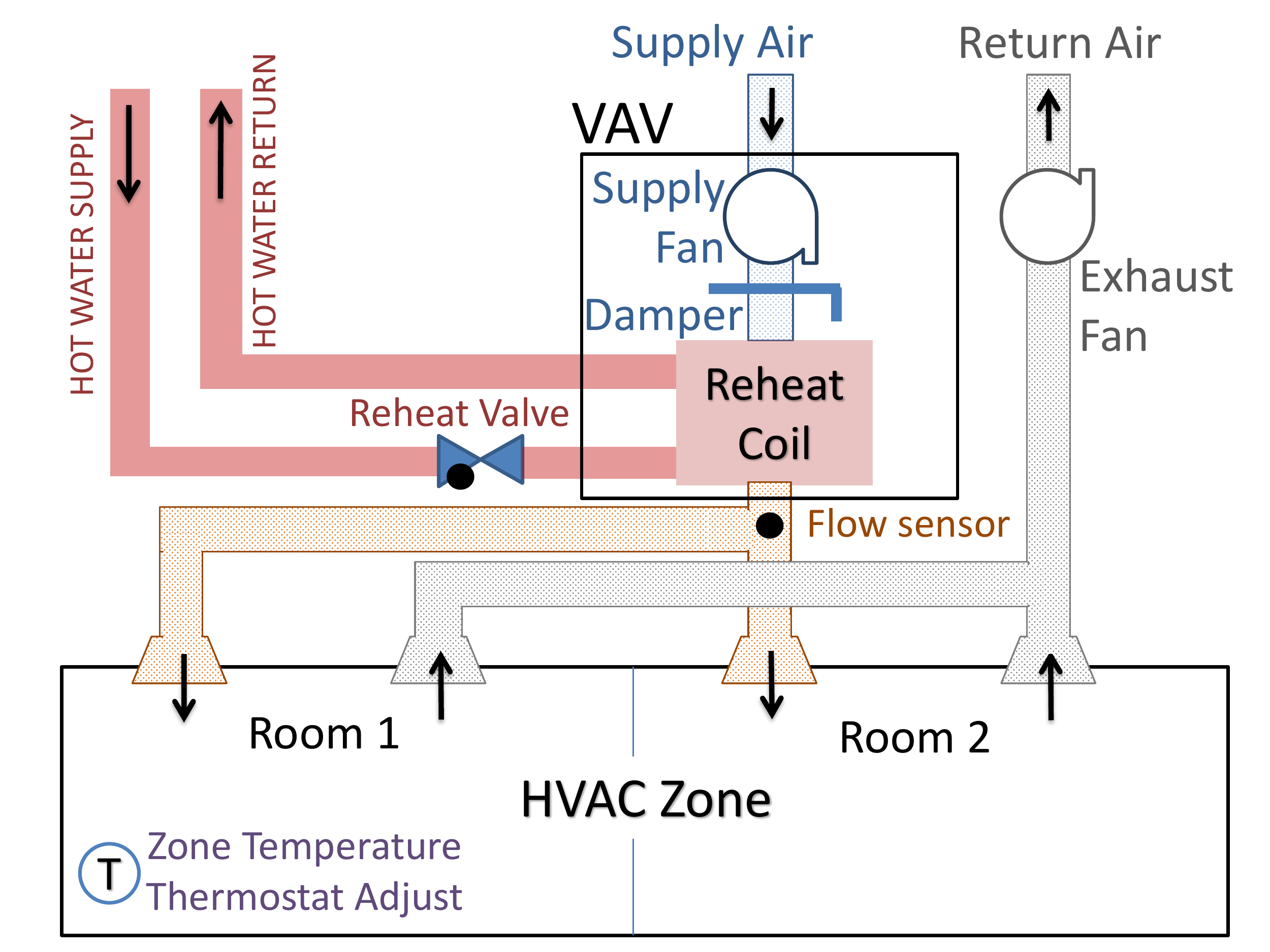}
\caption{VAV with reheat system used for controlling the temperature and airflow of discharge air in each HVAC zone in the CSE building ~\protect\cite{balaji2013zonepac}.}
\label{fig:zone_hvac}
\vspace{-5mm}
\end{figure}

Temperature control in buildings can be achieved by radiant heating systems, window air conditioners, packaged terminal units, etc. Our building uses Variable Air Volume (VAV) boxes that allow local temperature control, which is estimated to cover 20\% of cooling systems and are commonplace since 1990s~\cite{hydeman2003advanced}. For the remainder of this paper we will base our assumptions on buildings with similar controls given that new buildings and retrofits are all based on VAVs. VAVs allow each thermal zone to maintain its own thermal environment by modulating the amount of (cool) airflow in the zone using a damper and reheats the supply air (hot) when necessary. Figure~\ref{fig:zone_hvac} illustrates how VAV boxes manage temperature control of each thermal zone in the building. Supply and exhaust fans facilitate airflow for some zones, while the Air Handler Unit (AHU) determines the temperature of supply air depending on the cooling demand of the whole building. The temperature settings of the thermostats represents the only feedback for the control system. As there is only one thermostat installed per zone even if the zone encompasses multiple offices, spaces without thermostats, i.e. Room 2 in Fig.~\ref{fig:zone_hvac}, cannot provide direct feedback to the HVAC system. Hence, if an occupant in Room 2 is present during night/weekends, they cannot engage the HVAC system by pressing the thermostat power button in Room 1. Further, if Room 1 has high cooling demands, due for example to usage of heat dissipating equipment such as computers or copiers, Room 2 will be excessively cooled.

\section{Genie Design and Implementation}

\begin{figure*}[t!]
\includegraphics[width=\linewidth]{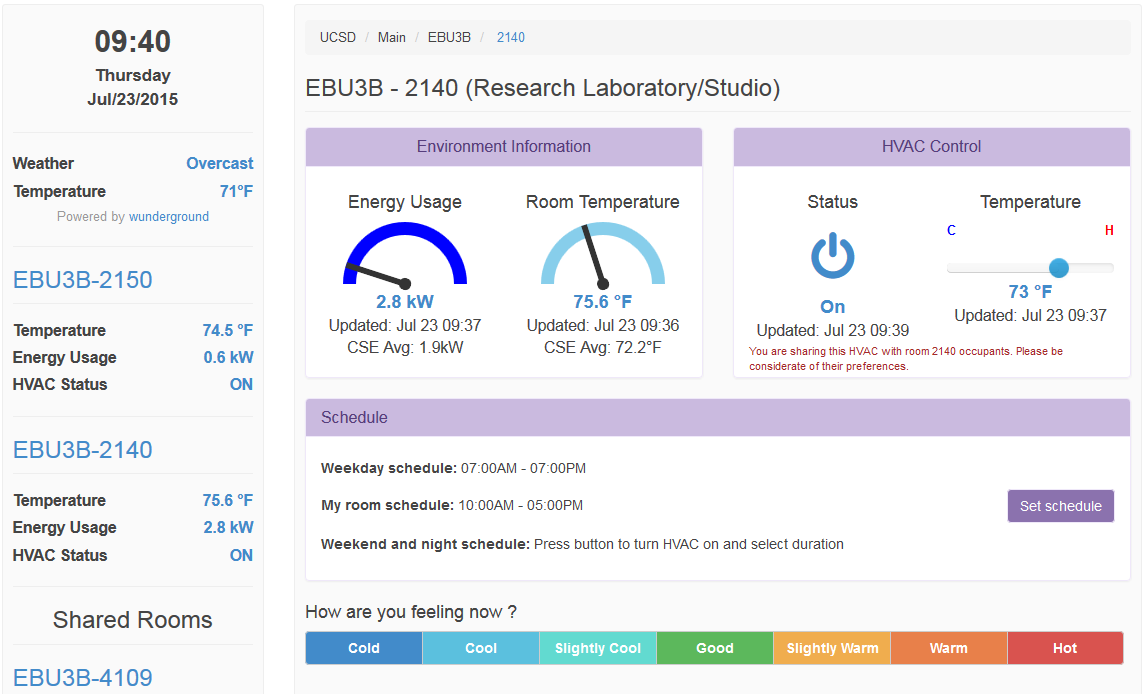}
\vspace{-5mm}
\caption{Screenshot of the Genie user interface. Users are given access to the rooms they have physical access to. They can change the temperature setpoint by $\pm3^{\circ}$F, choose to turn HVAC On/Off and set their own schedule.}
\label{fig:genie_screenshot}
\vspace{-4mm}
\end{figure*}

We designed Genie to mitigate many of the problems associated with the use of thermostats outlined earlier, and satisfy several design goals. First, we want thermostats to be more accessible and intuitive to use with occupants getting more control of their environment. Second, occupants should be able to send feedback to the building manager when needed. Third, we wanted energy conscious occupants to be able to get immediate feedback on the impact of their settings on the HVAC energy usage. Finally, for the particularly curious occupants we wanted to provide detailed data for the different sensors in the HVAC system.\\[-0.5cm]


\subsection{User Interface Design}
While designing the UI of Genie (Fig.~\ref{fig:genie_screenshot}), we emphasized transparent access to the HVAC data and functionality such as the current zone temperature, the temperature setpoint, HVAC system status, the estimated power consumed by the zone, as well as temperature control. We estimate zonal power consumption using available sensor data and heat transfer equations~\cite{balaji2013zonepac}. The Genie UI also shows a comparison of the current zone's temperature and power usage, with the average measurements of the overall building. Finally, we show the ``Last update time'' depicting the most recent change to the temperature, as a measure of the responsiveness of the system to changes made by occupants. 

Genie's web-based UI allows users to modify the temperature setpoint of their zone by $\pm3^{\circ}$F. Wyon et al.~\cite{wyon1996individual} show that this range is sufficient to meet the requirements of all the occupants in the building. To mitigate issues caused by multiple rooms sharing a single zone-level thermostat, we list the rooms belonging to the particular thermal zone in the UI while nudging occupants to be considerate with colleagues in the same zone. If a conflict of temperature preferences occurs, we suggest that occupants resolve this offline as the offices in the same zone are usually co-located. Users can set their own schedule, and the union of all the user schedules in a thermal zone is computed to be the zone schedule (default schedule is set to 7am - 7pm based on our experience). Users need to manually turn On the HVAC on weekends and set the number of hours they expect to be in their office through the UI, which puts that zone to the Occupied mode for the entire duration. If users explicitly turn Off the HVAC, we put that zone in Standby mode during weekday works hours, and in Unoccupied mode for nights/weekends. 

As shown in Fig.\ref{fig:genie_screenshot}, users can select different rooms using the navigation bar. They can request access to the rooms they have physical access to, which is manually verified before being approved. Genie only takes control of thermal zones whose occupants have registered, while the rest of the zones are managed by the traditional system. Note that the physical thermostat remains operational in zones with Genie controlling them, allowing users to manipulate temperature using either system. Public spaces such as kitchenettes, lobbies, and classrooms can in theory be accessed by any building occupant, which could lead to conflicts and abuse if anyone can exercise control. Hence, we initially restricted Genie access to only the personal offices in the building and then extended read-only access to public spaces a year later. In that way users could send feedback for public spaces to the building manager, who could decide to take action. 

In addition to real-time monitoring, control and feedback features, Genie also provides information to users who want to learn more or diagnose faults when they occur. Each of the sensor measurements -- airflow, temperature band, status of damper, etc. -- can be clicked to get historical values in the ``Show more details'' section. The navigation bar also provides weather information which has been shown to be useful~\cite{meier2011thermostat}. The About page illustrates the HVAC system functionality with detailed diagrams similar to the one in Fig.~\ref{fig:zone_hvac}.

\subsection{System Implementation}
\begin{figure}[t!]
	\includegraphics[width=\linewidth]{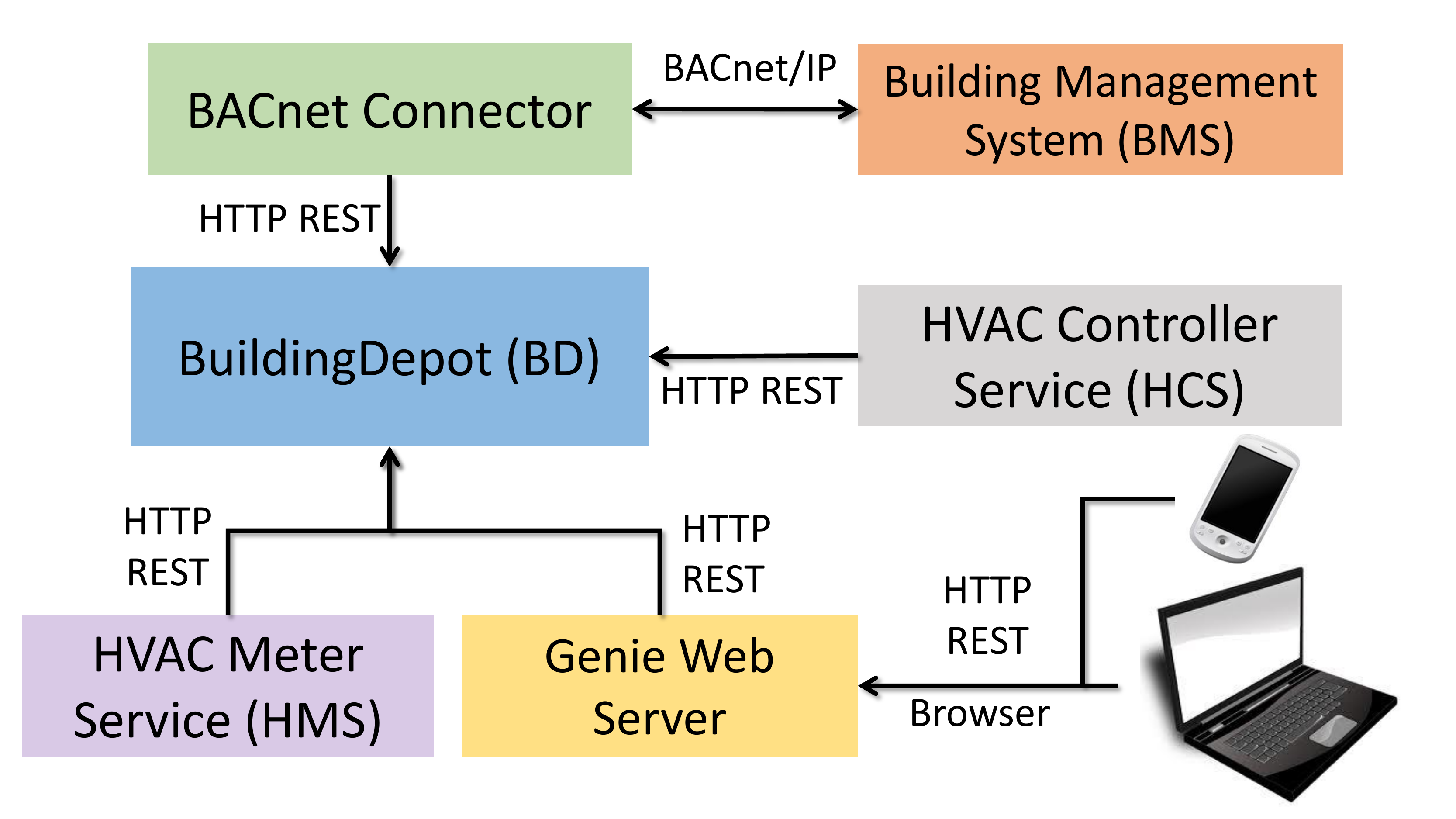}
	\caption{System architecture of the Genie web service. Data is collected from the BMS and stored in a building \emph{datastore}. Genie visualizes the data and provides RESTful APIs for browser access~\protect\cite{balaji2013zonepac}.}
	\label{fig:system_arch}
	\vspace{-5mm}
\end{figure}

 Thermostats in our testbed building are networked together and report data to a central server. The Building Management System (BMS), used by maintenance personnel, has a UI to view thermostat and other HVAC equipment data as well as to change configuration settings. Figure~\ref{fig:system_arch} shows the general architecture of \emph{Genie Web Service} the component that Genie uses to interface with the underlying BMS. We collect data from the BMS and store it in \emph{BuildingDepot}~\cite{agarwal2012buildingdepot} -- a RESTful web service developed specifically for building data management. BuildingDepot also enables control of the HVAC system, and exposes RESTful APIs for third party applications such as our Genie web service which provides the occupant facing services described above. A separate HVAC Meter Service (HMS) estimates zone-level power consumption~\cite{balaji2013zonepac} and the HVAC Controller Service (HCS) allows third party applications to control the HVAC without damaging the equipment. While users mainly access Genie via web browsers, it also exposes RESTful APIs for native smartphone applications as well as third party services. We use the Django framework,\footnote{http://www.djangoproject.com} to serve the Genie web application which is implemented on top of Bootstrap\footnote{http://getbootstrap.com}, and Flask\footnote{http://flask.pocoo.org} to deploy BuildingDepot. Both web services follow the Model-View-Controller architecture.

\section{Genie Deployment}
\label{sec:usage}
 
We announced Genie to all the occupants of our testbed building on October 15, 2013 over email. After the initial announcement, we created an internal mailing list for registered users. Three additional emails were sent to occupants to announce new features over the \nMonth month period. Users were not prompted in any other way to use this service. As of June 2015, there are \nUser registered users with a large number of these users being familiar with technology since they are student, staff and faculty in Computer Science. 

In addition to collecting logs and sensor data, we deployed a user survey and conducted interviews with occupants at the end of our study to understand their perspective on Genie's use. Our questions focused on knowledge of thermostats, comfort, features that were useful, effect of energy feedback and improvements that can be made to the system. 

In the remainder of this paper we present our mixed-methods analysis based on sensor data and log files collected by Genie from October 2013 to June 2015, combined with qualitative data from \nSurvey survey respondents and \nInterview contextual interviews. We anonymized data about users and the individual rooms to protect users' privacy as per our university's human research protection office's guidelines and our IRB approved study. 

\section{Longitudinal Study}
In our longitudinal analysis of thermostats' and Genie's use we focus on offices with individual occupants, and ignore common spaces such as conference rooms and kitchens. Individual offices make up 152 of the thermal zones in our building, of which 82 zones are controlled by Genie and the physical thermostat while the rest (70) are controlled by physical thermostat alone.

\begin{figure*}[ht]
\includegraphics[width=\linewidth]{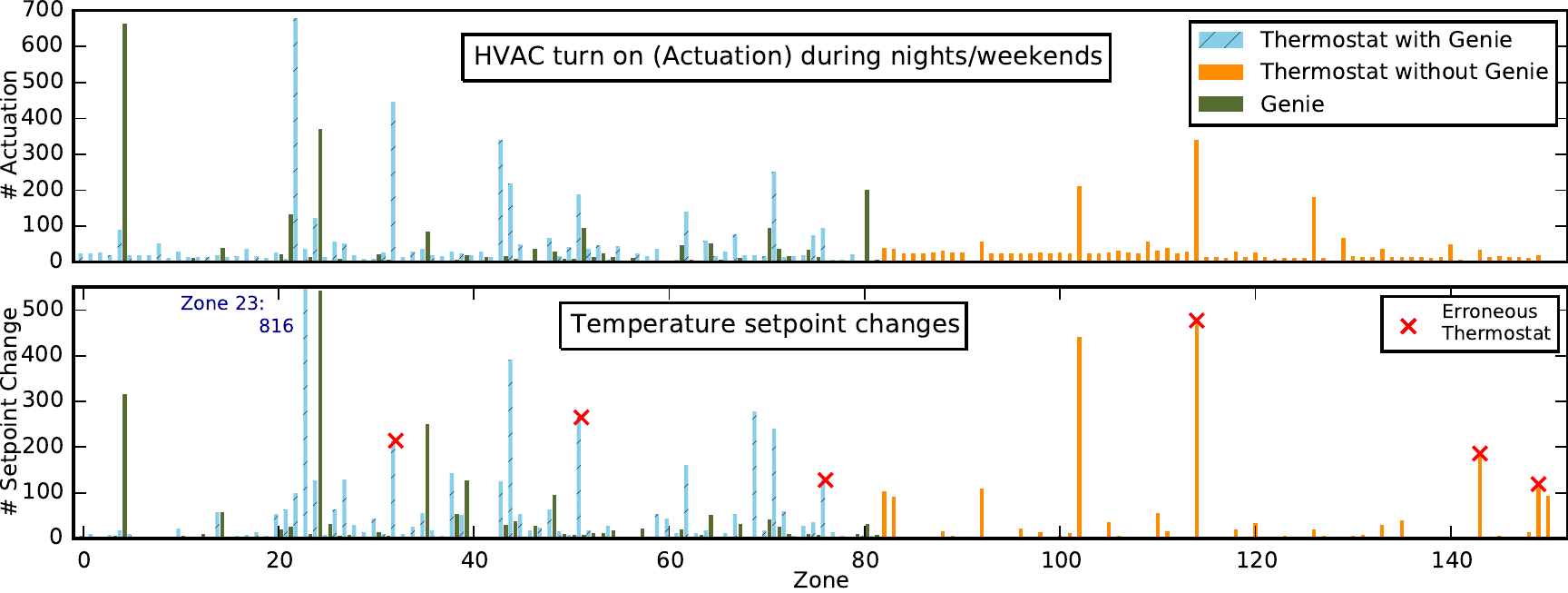}
\caption{Comparison of temperature setpoint changes and actuation during nights/weekends made using Genie and physical thermostats across 152 office thermal zones. Note that both the thermostat and Genie are used for the first 82 zones, and only the thermostat is used in the last 70 zones.}
\label{fig:usage_setpoint_zone}
\vspace{-3mm}
\end{figure*}


In order to compare usage and investigate emerging patterns we start by focusing our analysis on two main features provided by both the physical thermostat and Genie: (1) change of temperature setpoint and (2) HVAC actuation during nights (7pm - 7am) and weekends. Figure~\ref{fig:usage_setpoint_zone} shows an overview of the usage of Genie and the thermostats across all office zones. In general, thermostats are used much more than Genie, with thermostat usage constituting 73\% of all activity. However, in general, only a few zones show high activity, with 81\% of zones showing \textless 5 interactions with the system per month. To better understand how this overall usage is reflected in the two different interfaces we further analyze users' behavior by breaking it down in Physical Thermostat and Genie usage.

\subsection{Physical Thermostat}
Given the proliferation of thermostats in modern homes and buildings, it is not surprising that occupants used their thermostats at least a few times over our \nMonth month study. In fact, 74\% of our survey and interview participants knew about the use of the physical thermostat's slider to adjust temperature, and 36\% about the actuation button for nights/weekends. 

\subsubsection{Erroneous Thermostats}
\begin{figure}[hb]
\vspace{-4mm}
\includegraphics[width=\linewidth]{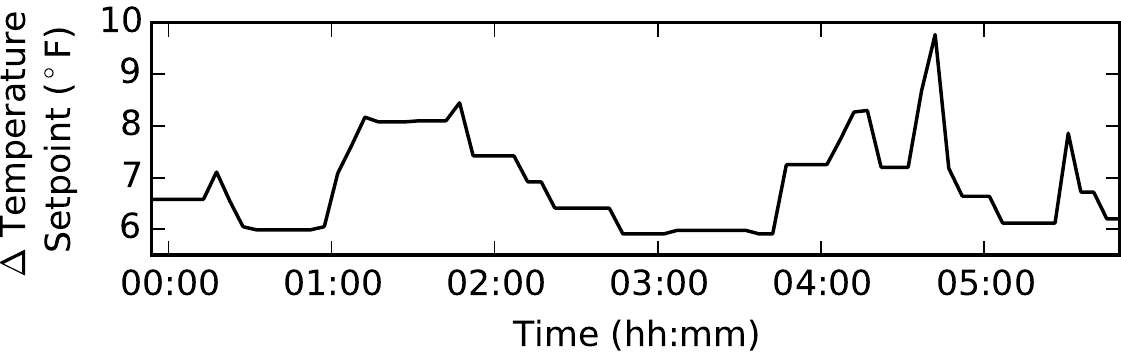}
\caption{An example of erroneous thermostat behaviour where changes occur frequently in the middle of the night. These changes are frequent in identified erroneous thermostats.}
\label{fig:erroneous_thermostat}
\end{figure}

Upon manual inspection of thermostat setpoint changes we observed that some of these changes were erroneously attributed to user interactions. Figure~\ref{fig:erroneous_thermostat} shows an example of frequent thermostat setpoint deviation in the middle of the night. Another thermostat showed an impossible change of $+12^oF$. These setpoint changes not only cause discomfort but also lead to energy wastage and equipment damage. We mark these thermostats as erroneous and do not consider them for further analysis. We consider a setpoint change only when it exceeds one-tenth the maximum range, i.e., for a thermostat slider with a range $\pm1^oF$, we consider a change of $\geq0.2^oF$. 

\vspace{8mm}

\subsubsection{Thermostats with High Activity} 
Some of the occupants are familiar with the thermostat, as one of our interviewee who works regularly on weekends surmises: \emph{``I only interact with it on weekends, because I figure that's when the temperature control is shut down centrally. [...] at some point if I'm sitting still in the office for a long time and the detectors don't detect any motion I think it turns off automatically and it starts getting warmer. I have to occasionally turn it on again.''} In reality, the HVAC is not connected to the sensors and turns Off after two hours independent of any motion, but the occupant knew to push the button repeatedly to keep HVAC working. We also found that occupants who work on weekends figured out how to use thermostats over time. As another interviewee explains: \emph{``I didn't even know you could push the button to turn on the AC at that time. So I would remember like... when I would come in on the weekends it would be hot and I wouldn't know what to do about it. [...] it wasn't until later when someone showed me how to use the thermostat and where it was even.''}

Upon manual inspection of data from zones which have high usage, we noticed that occupants in these zones have a habit of using the thermostat as soon as they enter the office in the morning or when it starts getting hot later in the afternoon. We saw an interesting correlation across users of thermostats with high activity in our data: in all of the cases the temperature setpoint range was widened to be \textgreater $\pm1^oF$, and the average range was $\pm7.3^oF$. Zone 23 with an abnormally high setpoint changes was a special case. Two of the occupants in a shared office had conflicting temperature requirements, and they changed the temperature settings several times in a day.

\subsubsection{Temperature Control and Discomfort}
Our interviews revealed that occupants have many misconceptions over how to use the thermostats and how it affected their office temperature. Many participants assumed the thermostat did not work, as an interviewee states: \emph{``I never thought it ever did anything. On the days it was too cold it stayed too cold.''} One of the occupants expressed frustration over the thermostat: \emph{``we didn't realize you had to actually push the button. I mean we were just pushing everything...''}, and as a result improvised their own solution: \emph{``Because it just blows down on me so forcefully that I actually went on top of my desk and I taped a manila folder to my ceiling.''} Use of space heaters (even in summer) is also a common solution used by occupants to combat overcooling by HVAC. Such improvisations not only cause excessive energy waste, but also leads to equipment damage. Occupants who did not have a thermostat in their offices often did not realize they had control over the temperature. As another interviewee states: \emph{``I was freezing to death. You can shut the door if that helps. I was freezing to death and I didn't know where the thermostat was to make at least my area...at least comfortable for me...''}. Our surveys corroborate these findings reporting an average comfort level of 2.9 out of 5 with the use of thermostats.\\[-.5cm] 

%
%
%
%


%
%
%
%

\subsection{Genie}

After looking at our log files we discovered that the overall usage of Genie seemed to be much lower than the thermostats (see Fig.~\ref{fig:usage_setpoint_zone}). However, after carefully considering the possible reasons behind this potentially disappointing result, we recognized that Genie allows for a wider temperature control than thermostats, which may result in reduced number of changes as occupants are comfortable with that temperature. Furthermore, the physical thermostat turns On the HVAC only for 2 hours at a time, while Genie expands that to up to 14 hours. Thus, it is possible that Genie's absolute actuations count does not correspond to effective usage of the interface. Moreover, our survey indicated that comfort level after using Genie increased to 4.2 out of 5 vs 2.9 using thermostats, with the difference being statistically significant ($F_{1,33}=29.42,p=<0.0005$). To investigate how users consistently used temperature control across \nMonth months and why they reported such an increased comfort level, we further analyzed Genie's logs.


\subsubsection{Engagement over time}
\begin{figure}[t!]
	\centering
	\includegraphics[width=.95\linewidth]{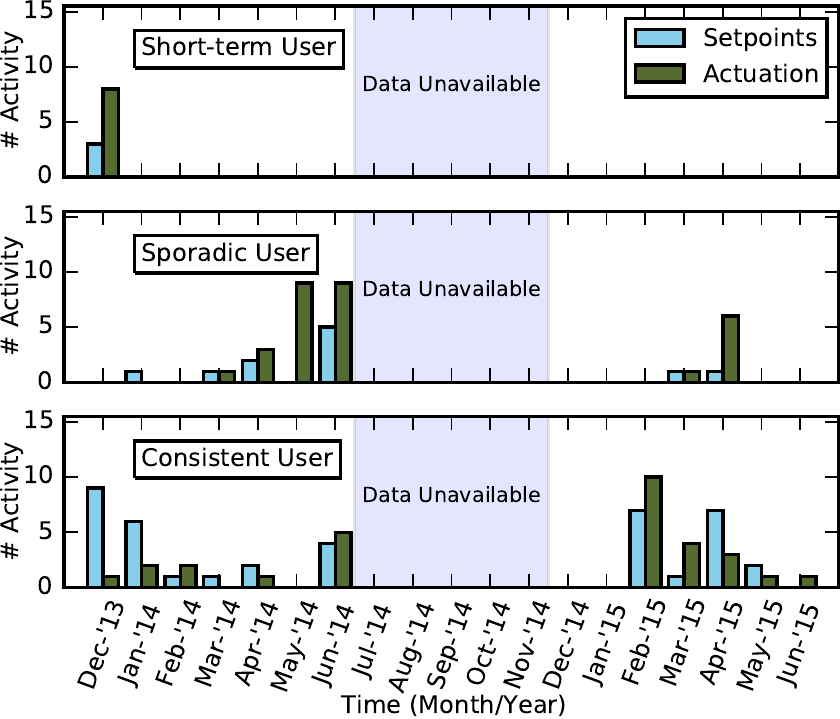}
	\caption{Genie activity comparison for a representative user from each category.}
	\label{fig:genie_all_users}
	\vspace{-5mm}
\end{figure}

Although Genie logs were only available for 122 of \nUser users and for 13 out of the \nMonth months of deployment (logs are not available for the initial two months and for six additional months as indicated in Fig.~\ref{fig:genie_all_users}) we were still able to get a detailed view of Genie's usage characteristics. Based on this analysis we were able to categorize Genie's users into four distinct types:\\[-0.5cm]

\begin{itemize}
	\item \textbf{One-time}: Users visit the page a few times after registration and do not visit again.\\[-0.6cm]
	\item \textbf{Short-term}: Users actively use Genie for $\leq2$ months.\\[-0.6cm]
	\item \textbf{Sporadic}: Users whose regular use of Genie is spread across more than 2 months, although interspersed with gaps in their usage for several months.\\[-0.6cm]
	\item \textbf{Consistent}: Users who used Genie consistently for more than 6 months.
\end{itemize}

Figure \ref{fig:genie_all_users} shows usage data from logs for three example users from each category and Table \ref{tab:user_categories} summarizes the results across all users.  From our analysis we conclude that a significant portion (45.1\%) of users were actively engaged in using Genie for more than two months after their registration. 

We investigated further to find out the specifics of when and why people wanted to use Genie through our surveys and interviews. Our data revealed that Genie was especially useful when users did not have a thermostat in their office. As one interviewee explains: \emph{``I didn't actually use the older thermostat because I don't have a thermostat in this room. ... for me Genie is great because I have personalized access to my room.''} Users also liked the precision of control made available by Genie, as one survey respondent comments: \emph{``Digital control of the temperature is very, very useful. Moving the slider [on the thermostat] still leaves a lot of uncertainty as to what exactly will happen, and the temperature setting helps.''} One of the survey respondents commented on how temperature control affected his productivity: \emph{``Genie is awesome and has made a real difference in my ability to work in my office. I get migraines that are correlated with higher temperatures, and Genie allows me to set the office temperature to 67, which greatly reduces occurrence.''}.

\begin{table}[t!]
	\centering
	\begin{tabular}{r c c c c}
		\toprule
		{\small\textbf{User Types}}    & One-time  & Short-term & Sporadic & Consistent\\
		\midrule
		\small{\textbf{\% Users}} & 24.6\% & 30.3\% & 23.8\% & 21.3\% \\
		\bottomrule
	\end{tabular}
	\caption{Percentage of Genie users per category: one-time, short-term (\textless 2 months), sporadic (gaps in usage) and consistent ($>$ 6 months). 
	}
	\label{tab:user_categories}
	\vspace{-5mm}
\end{table}

For the \emph{consistent} users we found that Genie is often actively used because offices are uncomfortable on a regular basis. As one user says: \emph{``I generally think its fine ... only in the late afternoon I have to make it cooler''}. On the other hand \emph{sporadic} users use Genie occasionally because offices are already quite comfortable:, as reported by one of the interviewees: \emph{``I mean, I haven't used it a lot. I just...uhm...will change the temperature if it's like too hot or too cold. And on the weekends if I'm working here I'll turn it on because the AC doesn't turn on automatically.''}. \emph{Short-term} users often indicated how the initial interest was high and then it vanished with time: \emph{``I used it frequently at some point as in usually over the weekend, I would tweak the temperature through the web interface. Then nowadays I don't come in as often in the weekends. So if I do come, I might set up the thermostat manually coming in the room. Then usually I don't have to deal with it until I leave...so yeah, I may not have been used the web interface for a while now.''}. Finally, \emph{one-time} users typically forget the URL, or the password for their account, and do not visit the web page after their initial registration. As one user indicated: \emph{``It looks pretty friendly. It's more of a matter of out-of-sight...out-of-mind.''}

\subsubsection{Dual Thermostat Usage}
Many of our survey respondents revealed they used the physical thermostat despite having a Genie account. One of the main reasons echoed by several users was that the thermostat was sometimes easier to access compared to opening the computer and controlling the temperature via the web app. As one user says: \emph{``I don't have to pull up the web interface. It's just a dedicated slider on the wall, which is pretty easy for occasional tweaks.''} Another reason for using the physical thermostats was that many occupants were confused about the relationship between Genie and the thermostat on the wall. As one survey respondent explains: \emph{``I don't quite understand how the physical thermostat and Genie interact and so I often adjust both.''} Both Genie and the thermostat were functional, but Genie does not directly reflect the changes made through the thermostat slider. Having access to both controls confused some users; we realized that this is a design flaw and we are planning to address that in our future work, with the Genie interface directly reflecting the physical thermostat changes. 


\vspace{8mm}
\subsubsection{Thermal Feedback from Occupants}
Genie introduced the ability to send feedback on how comfortable occupants are in their offices. Some users were unclear on the utility of the feedback, and whether it affected their HVAC settings. Users therefore initially sent feedback to express their comfort level or justify their control actions. As one of the feedbacks said: \emph{``Felt cool for the past 1-2 wks. Just tried changing the room temp from 73 to 75 hoping we feel a difference!''} Other users would ask questions about the interface: \emph{``AC seems to be off during weekend. Can I/anyone turn it on?''} Many users initially sent \emph{``Good''} feedback, which we interpreted as being satisfied with the HVAC system. However, the majority of feedback messages we received were linked to users being uncomfortable despite changing their temperature settings, or complaining about Genie or the HVAC system not working correctly. 

\begin{figure}[t!]
\centering
	\includegraphics[width=\linewidth]{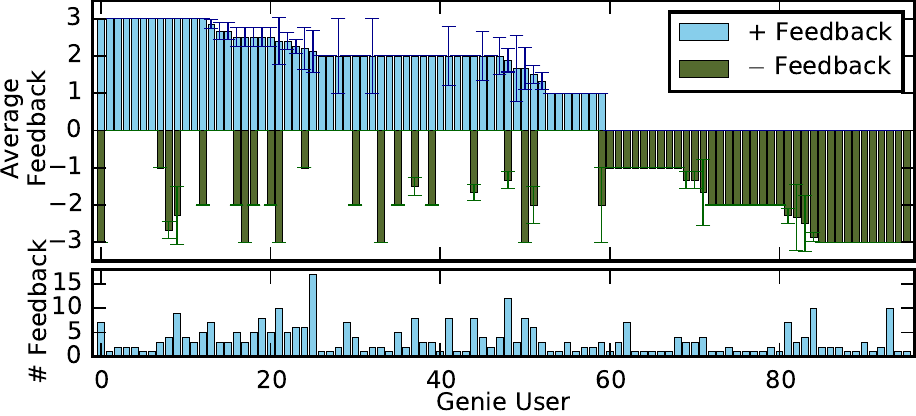}
	\caption{Distribution of feedback given using Genie across all users on standard PMV 7-point scale. Only 106 out of 220 users provide feedback, and 36 of those users only provide feedback once. Feedbacks help identify extreme conditions in the offices and insights into HVAC faults.}
	\label{fig:genie_feedback}
  \vspace{-5mm}
\end{figure}

Occupants' feedback also served an additional means. As facilities managers do not have time to inspect the problems in every room in a building, faults that occur at the office zone level are often ignored and remain undiscovered unless an occupant sends a complaint~\cite{teraoka2014buildingsherlock}. The feedbacks from Genie proved to be a valuable resource to identify such faults and correct them to improve occupant comfort. Figure~\ref{fig:genie_feedback} shows the distribution of comfort feedbacks sent by the users along with the mean values of their comfort level in the standard 7 point scale. As can be seen from the graph, most users only send a few feedback messages. These messages usually correspond to extreme discomfort levels. The textual feedback sometimes elaborates on the issue. For instance, one user comments: \emph{``I am wearing a sweater but I am cold in the office. Walking in the corridor, I am much colder. My hands are really cold.''} Sometimes the users will directly send a symptom of a fault, for example: \emph{``It's 64 in here now, though the setting is the max allowed at 73.''} During our deployment these messages allowed the building manager to uncover many unknown or unreported system faults. Examples include sensors which stopped reporting information, thermostats which were blocked by computers, dampers getting stuck, Genie not reporting data, etc. 

\subsubsection{Energy Feedback to Occupants}
\begin{figure}[t!]
\centering
	\includegraphics[width=.95\linewidth]{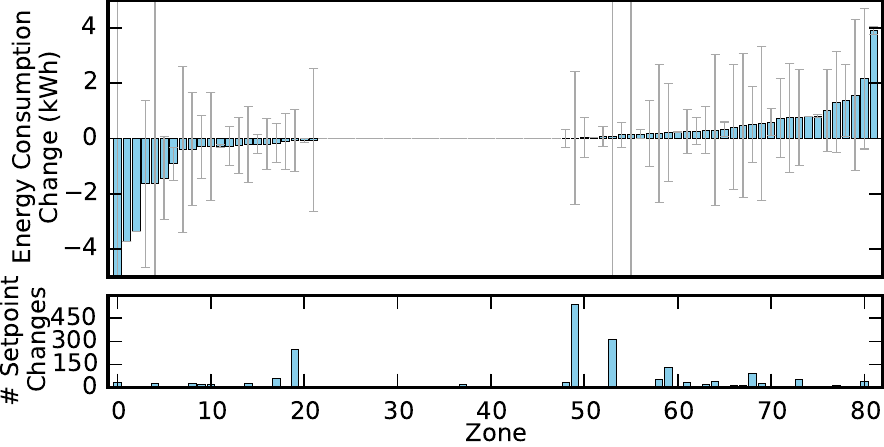}
	\vspace{-1mm}
	\caption{Average energy consumption difference 2 hours before and after a change in setpoint by a Genie user across all zones.}
	\label{fig:energy_difference}
\vspace{-5mm}
\end{figure}

Genie provides the estimated energy consumed within the thermal zone to the users and a normalized average energy consumption for the building to allow users to compare their energy usage with other zones in the building. While prior work did show the effect of energy feedback on occupant's behavior (5\% reduction), the results were preliminary with a small set of users and over 5 days~\cite{balaji2013zonepac}. As part of our study we analyzed the effect of energy feedback across 21 months. Figure~\ref{fig:energy_difference} shows the effect on zone energy consumption due to a temperature setting change by the user. We compare the energy consumption two hours before and after a change made by the user to infer if the user made an energy conscious decision. The data shows that the energy consumption could equally decrease or increase, and there is no bias towards energy conserving settings. As we show later, Genie zones show a 3\% decrease in energy consumption on weekdays and 31\% increase in weekends compared to physical thermostat zones.


In addition to the effects we registered in our system, we investigated the personal occupant's perception in terms of added energy consciousness. Our survey revealed that users were divided on whether they were more energy conscious after using Genie, with a mean score of 2.8/5. Many users commented that their comfort was a clear priority over the energy consumption. As one interviewee states: \emph{``If I'm hot dude...I'm going to turn it on. I mean uh...I got work to do. You know...if I got to use a little bit of wattage I don't care.''} Some users agree that it is good to be aware of the energy consumption, but it does not change their behavior in any way. As one user comments: \emph{``I do care, but admittedly would do whatever I needed to be comfortable without regard to energy consumption.''} A subset of users, however, expressed a desire to better understand their energy footprint, and wanted more indication in the interface on how they could act upon decreasing it. One user states: \emph{``I think it would be helpful even to see what your peers...what their energy consumption is. Just to kind of see if I'm conserving a lot more, or...wow...I'm way over the top. Maybe I need to start being more conscientious about things.''} 

38\% of the users responded that they were more energy conscious with the feedback Genie gave them. Therefore, although many users do not care, energy consumption's feedback does have an overall impact in behavior on an important subset of our user base.

\subsubsection{Genie's Limitations}
Despite the overall positive feedback from our users, Genie introduced its own set of problems and exposed some limitations. A common issue among many users was that the HVAC control was limited to once every 10 minutes. This was our design decision to protect the HVAC equipment from excessive usage. As a consequence of this conservative setting Genie was unresponsive to some specific user's behaviors and intended interactions with the system. For instance, when users made a minor mistake with the temperature setting, or accidentally pressed a button, the system would not let them change the settings for the next 10 minutes. As one user explains: \emph{``I was trying to adjust it and I moved it down and I slipped...and so I let go of the mouse and it only moved a half degree. Then it was like you can do this again in 10 minutes...''} Another major issue occurred when Genie was temporarily unavailable due to system updates or maintenance. We have had only a few instances which led to unavailability over some weekends, and at that time users had to revert to using thermostats. One user sent us a message when Genie was down: ``For some reason the A/C wasn't running ... I don't have a thermostat in my office (it's in another office next to mine that I don't have access to), so genie was my only hope''. Hence, when Genie fails, an alternative such as manual thermostat override should be available. This is important in case occupants cannot access a networked device or in case of a software failure. Thus, the system needs to be carefully designed to address these scenarios.

\subsubsection{Additional Features}
Genie provides several other features, most of which remain unused. Most users do not set their personal schedule if the default schedule is enough to make them comfortable. The history of each sensor can be obtained by clicking on the measurement (e.g. $72^{\circ}$F) in the UI. Although many users indicated history was useful, they did not realize this feature was available. We provide detailed sensor data and details about what each sensor means, but this is almost never used. 

We did not provide users access to shared spaces such as conference rooms and lobbies due to conflicts that may occur between requests. To extend Genie functionality we synchronized the online conference room calendar with the Genie schedule so that users have control over the HVAC settings for the duration of the meeting. The HVAC is turned down during non-meeting times to save energy. Although many users liked this feature when we announced it, most users either forgot about it or did not eventually use it. 

\section{Impact On the HVAC System}
As Genie provides more flexibility for occupants to control their temperature and turn HVAC On/Off, one of the risks from a building manager perspective is that Genie could lead to an increase in overall energy consumption or deviation of operation from the HVAC managament's original design and intended purpose. To investigate the impact of this added flexibility, we compared the overall energy consumption and the extent of control exercised using Genie versus the physical thermostats.

\subsubsection{Energy Consumption}
\begin{figure}[t!]
	\includegraphics[width=\linewidth]{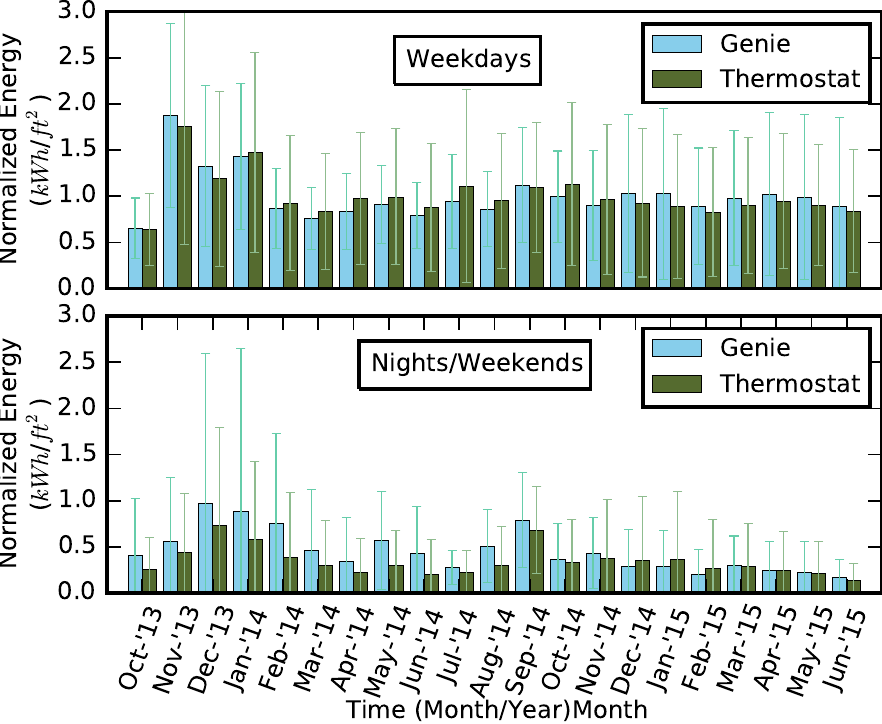}
	\caption{Comparison of Genie and Thermostat zone energy consumption across \nMonth months. The energy consumption has been normalized by area to account for varying room sizes. Other confounding factors such as presence of windows is assumed to be randomly distributed.}
	\label{fig:energy_compare}
	\vspace{-5mm}
\end{figure}

We first focus our attention on how Genie impacted energy consumption. Figure~\ref{fig:energy_compare} shows a comparison of normalized energy consumption for weekdays and weekends separately. The weekday graph indicates that the energy consumption of Genie zones is comparable to the thermostats, and overall Genie zones save 3.5\% energy, The difference is statistically insignificant ($F_{1,70}=0.001,p=ns$), but we can still confidently say that Genie's usage is not linked too more energy consumption during the week. On the weekends, Genie zones consume more energy on average, and this points to the fact that users utilize Genie regularly to actuate the HVAC on weekends. Hence, this excess in energy consumption is justified as it serves to keep the occupants comfortable. Genie zones consume 31.6\% more energy than zones with thermostats during the weekends but the difference is statistically insignificant ($F_{1,70}=2.59,p=0.11$). Comparing the overall energy consumption considering both weekends and weekdays, Genie zones consume 3.4\% more than thermostat zones, but it is again statistically insignificant ($F_{1,70}=0.092,p=ns$). Therefore, long term use of Genie has not had a significant effect on HVAC energy use. 

\subsubsection{Temperature Swing}
As the temperature setting can be changed up to $6^{\circ}$F in Genie, users may tend to change the temperature settings to its extremes which may lead to excessive energy consumption or large swings in airflow.  We compared the deviation in temperature settings across different zones over \nMonth months. Surprisingly, some physical thermostats show more deviation than Genie, with up to $6^{\circ}$F standard deviation. This can be attributed to those physical thermostats whose range have been increased by the building manager in response to comfort complaints. The occupants do not know by how much they are changing the temperature as there is no indication in the thermostat. There are a total of 63 out of 152 thermostats whose range is larger than the designed $\pm1^{\circ}$F, and the building manager does not keep a track of these thermostat changes. On the other hand, despite having the freedom to change the temperature by $6^{\circ}$F in Genie, surprisingly most extreme changes in Genie are around the $4^{\circ}$F mark. The standard deviation for the change is $\pm2.0^{\circ}$F, compared to $\pm3.5^{\circ}$F in thermostats, and this difference is statistically significant ($F_{1,>>100}=95,p<0.0005$). All in all, we can see here that providing users with clear information and more control results in better overall behavior than providing a slider without information on the thermostat. 









\section{Lessons Learned and Design Guidelines}
Our combined analysis of thermostat's and Genie's usage data with user interviews and surveys revealed that the thermostats in our building fail to provide clear status and feedback information to occupants. In addition, some occupants do not know where thermostats are located, or do not have access to them. These findings confirm the outcomes of prior studies~\cite{karjalainen2009thermal,karjalainen2007user}. We showed here how software-augmented thermostats can alleviate these issues as well as provide additional features such as getting feedback from occupants. Systems like Genie are especially attractive for existing buildings, where retrofitting can cost from \$500-\$2,500 for each thermostat~\cite{retrofit-cost}. Based on our experience with the design and development of Genie and our longitudinal study, we discuss below six specific design guidelines that we believe will guide and inform the future design and development of software-augmented thermostats.

\subsubsection{Relationship to Physical Thermostats}
Software thermostat should not aim to replace the physical thermostat. Thermostats have been around since 1572~\cite{walker2008residential}, and many occupants are familiar with its basic functions. We claim that physical thermostats can still provide basic functions and occupants should be able to use them when they do not have access to a networked device or when there is a software failure. However, it is important that both the physical and software thermostats show a similar interface, and are synchronized with each other's updates, so that users do not get confused with the relationship between them.\\[-0.4cm]

\subsubsection{Clarity of Information}
Users value the precision of information available in a software graphical interface, since it allows them to better comprehend what the HVAC system is trying to accomplish. Thus, although a simplified interface is necessary, it should not leave out essential information such as if the HVAC is working currently, and what temperature settings are in use. Our data shows that users visit the software interface only when they feel uncomfortable, and that accurate information allows them to infer the current status quickly.\\[-0.5cm]

\subsubsection{Provide Adequate Control}
Users expressed immense satisfaction in having the ability to control their local office temperature, which confirms findings from prior studies~\cite{dear2013progress,paciuk1989role}. Showing users how much control is available to them and how it affects the HVAC operation allows users to make intelligent decisions. Our data shows that users are careful with their control decision and the impact on HVAC operation and energy consumption is minimal. 

\subsubsection{Comfort Complaints and Feedback to Managers}
Comfort feedback not only provides building managers information on the level of comfort of occupants, but also helps in identifying hard to detect faults such as thermostat blockage. Fault detection algorithms and control strategies can use this information to crowdsource comfort information and further tune the HVAC system as per user requirements.\\[-0.5cm]           

\subsubsection{Actionable Information on Energy Usage}
Many users like energy consumption feedback, and a number of them even indicated active interest in using the information to save energy. Prior studies have shown that providing actionable energy reduction information can be effective in residential settings~\cite{roberts2004consumer}. Users need similar information in offices, as one interviewee requested: \emph{``... if by changing this 1 degree I would save this percent of energy, I would do it.''}\\[-0.5cm]

\subsubsection{Prediction and Additional Features}
In a software interface, users expect fast reaction times to inputs. Thus, the system needs to hide HVAC latency and show the effective change that will occur later. Another strategy is to provide users with predictions of HVAC behavior due to a change in setting, which has shown to be effective in homes~\cite{sauer2009designing}. Further, features such as historical data should be intuitive to discover for users to actually use them.\\[-0.6cm]

\subsection{Limitations}
We note that our study of physical thermostats and Genie usage has been conducted in a university building located in a temperate climate zone in the US. The analog thermostat we studied is from Johnson Controls, a popular vendor who install HVAC systems across 125 countries. Although the thermostat model we consider is installed across most buildings in our university campus, it predates the latest digital model provided by the vendor. Therefore, more research is needed to verify our findings across different cultures, climate zones and types of thermostats. Finally, our occupants are all from a Computer Science building, and more research is required to generalize our findings to other population pools.

\section{Conclusion}
We designed a software augmented thermostat, called Genie, that provides pertinent HVAC status information to the occupants and enables adequate control over their local temperature. We introduced additional features such as comfort feedback from occupants as well as energy consumption information to increase occupant awareness. To evaluate Genie, we deployed it in a five floor university building, and studied its usage as compared to the physical thermostat alone over \nMonth months. 

We show that occupants have misconceptions about thermostat usage, and some of them did not know where thermostats were located. Genie users were satisfied with the clarity of information and level of control available, and 45\% of users showed longer term engagement with the system. In addition, comfort feedback from users provided insights into non-obvious HVAC faults. The energy feedback provided by Genie increased user awareness with a subset of the user base motivated to change their behavior. Based on our usage analysis and design experience, we outlined key design guidelines for software augmented thermostats. 

All in all, we believe that the insights presented in this study will benefit researchers and designers who want to further investigate temperature control in office buildings and develop user facing smart building applications

%
%
%
%
%
\balance{}
\newpage

\bibliographystyle{SIGCHI-Reference-Format}
\bibliography{sigproc}

\end{document}